\documentclass{iau}
\usepackage{amsmath,amssymb,bm,natbib,graphicx}

\newcommand{\vv}{\mathbf{v}}
\newcommand{\vB}{\mathbf{B}}
\newcommand{\vJ}{\mathbf{J}}

\newcommand{\bomega}{\boldsymbol{\Omega}}

\newcommand{\dvg}{\boldsymbol{\nabla}\!\bcdot\!}
\newcommand{\curl}{\boldsymbol{\nabla}\!\boldsymbol{\times}\!}

\newcommand{\cross}{\!\boldsymbol{\times}\!}

\editors{J.~J. Eldridge}
\pubyear{2017}
\volume{329}
\jname{The lives and death-throes of massive stars}

\title[Dynamo Scaling Relationships]{Dynamo Scaling Relationships}
\author[Augustson, Mathis, Brun, \& Toomre]{Kyle Augustson$^1$, St\'{e}phane Mathis$^{1}$, Sacha Brun$^{1}$,
  \and Juri Toomre$^2$}

\affiliation{$^1$Laboratoire AIM Paris-Saclay, CEA/DRF -- CNRS -- Universit\'{e} Paris Diderot, IRFU/SAp Centre de
  Saclay, F-91191 Gif-sur-Yvette Cedex, France\\ $^2$JILA, University of Colorado -- Boulder, Colorado, USA 80309}

\begin{document}

\maketitle

\begin{abstract}
  This paper provides a brief look at dynamo scaling relationships for the degree of equipartition between magnetic and
  kinetic energies. Two simple models are examined, where one that assumes magnetostrophy and another that includes the
  effects of inertia. These models are then compared to a suite of convective dynamo simulations of the convective core
  of a main-sequence B-type star and applied to its later evolutionary stages.
\end{abstract}

\section{Introduction}

The effects of astrophysical dynamos can be detected at the surface and in the environment of many magnetically-active
objects, such as stars \citep[e.g.,][]{christensen09,donati09,donati11,brun15}.  Yet predicting the nature of the
saturated state of such turbulent convective dynamos remains quite difficult. Nevertheless, one can attempt to
approximate the shifting nature of those dynamos. There may be the potential to identify a few regimes for which some
global-scale aspects of stellar dynamos might be estimated with only a knowledge of the basic parameters of the
system. For instance, consider how the magnetic energy of a system may change with a modified level of turbulence and
also how rotation may influence it. Establishing the global-parameter scalings of convective dynamos, particularly with
stellar mass and rotation rate, is useful given that they provide an order of magnitude approximation of the magnetic
field strengths generated within the convection zones of stars as they evolve from the pre-main-sequence to a terminal
phase. This could be especially useful in light of the recent evidence for magnetic fields within the cores of red
giants, pointing to the existence of a strong core dynamo being active in a large fraction of main-sequence,
intermediate-mass stars \citep{fuller15,cantiello16,stello16}. In turn, such estimates place constraints upon transport
processes, such as those for angular momentum. \vspace{-0.25truein}

\section{Scaling of Magnetic and Kinetic Energies} \label{sec:forcescaling}

Convective flows often possess distributions of length scales and speeds that are peaked near a single characteristic
value. One estimate of these quantities in stellar convection zones assumes that the energy containing flows possess a
kinetic energy proportional to the stellar luminosity ($L$) that is approximately
$\mathrm{v_{rms}} \propto \left(2L/\rho_{\mathrm{CZ}}\right)^{1/3}$ \citep{augustson12} , where $\rho_{\mathrm{CZ}}$ is
the average density in the convection zone. However, such a mixing-length velocity prescription only provides an order
of magnitude estimate \citep[e.g.,][]{landin10}. Since stars are often rotating fairly rapidly, their dynamos may reach
a quasi-magnetostrophic state wherein the Coriolis acceleration also plays a significant part in balancing the Lorentz
force.  Such a balance has been addressed and discussed at length in \citet{christensen10}, \citet{brun15}, and
\citet{augustson16}.

\begin{figure}[!t]
  \centering
  \includegraphics[width=\linewidth]{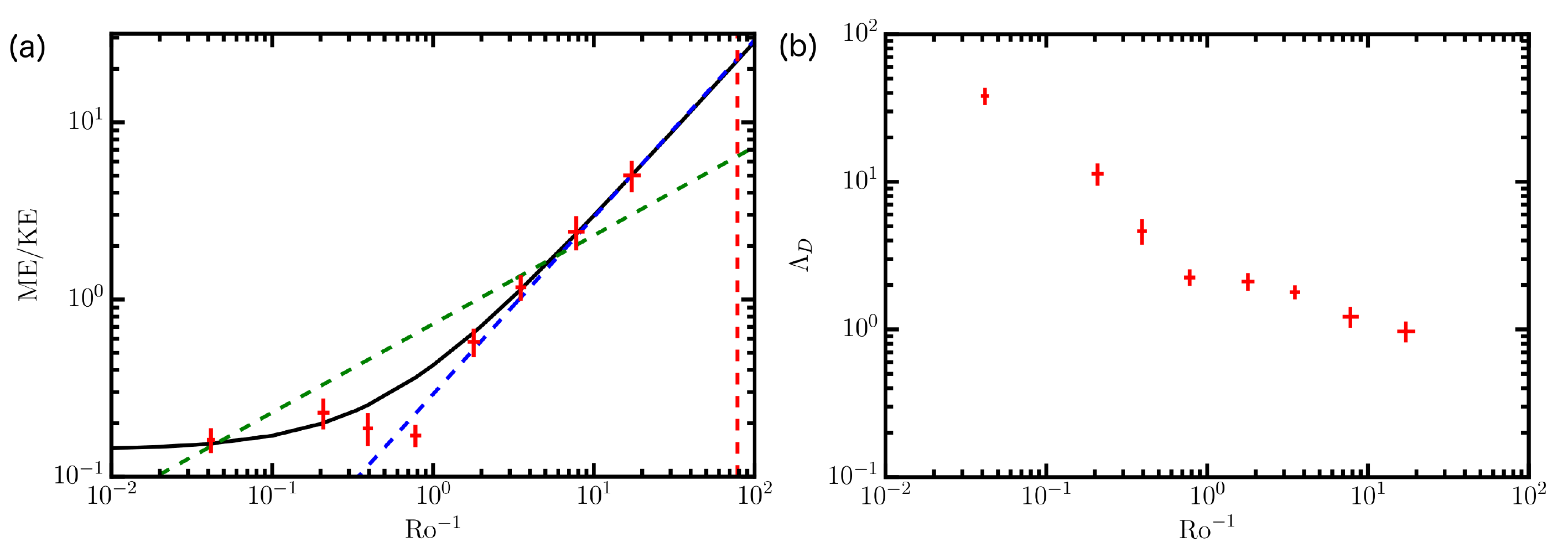}
  \caption{(a) The scaling of the ratio of magnetic to kinetic energy ($\mathrm{ME/KE}$), with data from
    \citet{augustson16}. The black curve indicates the scaling defined in Equation (\ref{eqn:forcescaling}), with
    $\beta=0.5$. The blue dashed line is for magnetostrophy ($\beta=0$).  The green dashed line represents the
    buoyancy-work-limited dynamo scaling, where $\mathrm{ME/KE}\propto \mathrm{Ro}^{-1/2}$. The red dashed line
    indicates the critical Rossby number of the star, corresponding to its rotational breakup velocity. (b) The scaling
    of the dynamic Elsasser number ($\Lambda_D$) with inverse Rossby number in simulations from \citet{augustson16}. The
    uncertainty of the measured Rossby number and energy ratio or dynamic Elsasser number that arises from temporal
    variations are indicated by the size of the cross for each data point.}
  \label{fig:scaling}
\end{figure}

In \citet{augustson17a}, it is shown that one can derive a scaling relationship based upon the vorticity equation.  In
particular, integrating the enstrophy equation and ignoring any loss of enstrophy through the boundary requires that

\vspace{-0.2truein}
\begin{center}
  \begin{align}
    \curl{\left[\rho\vv\cross\boldsymbol{\omega}+2\rho\vv\cross\bomega+\frac{\vJ\cross\vB}{c}+\dvg{\sigma}\right]}=0. \nonumber 
  \end{align}
\end{center}

\noindent Thus, the primary balance is between inertial, Coriolis, Lorentz, and viscous forces.  Scaling the derivatives
as the inverse of a characteristic length scale $\ell$ and taking fiducial values for the other parameters in the above
equation yields $\mathrm{ME/KE \propto 1 + Re^{-1} + Ro^{-1}}$, when divided through by
$\rho\mathrm{v_{rms}^2}/\ell^2$. Here the Reynolds number is taken to be $\mathrm{Re} = \mathrm{v_{rms}}\ell/\nu$.
However, the leading term of this scaling relationship is found to be less than unity, at least when assessed through
simulations.  Replacing it with a parameter to account for dynamos that are subequipartition leaves

\vspace{-0.2truein}
\begin{center}
  \begin{align}
    \mathrm{ME/KE} \propto \beta(\mathrm{Ro, Re}) + \mathrm{Ro}^{-1}. \label{eqn:forcescaling}
  \end{align}
\end{center}

\noindent Here $\beta$ is unknown apriori as it depends upon the intrinsic ability of the non-rotating system to
generate magnetic fields, which in turn depends upon the specific details of the system such as the boundary conditions
and geometry of the convection zone.

\begin{figure}[!t]
  \centering 
  \includegraphics[width=1.15\linewidth]{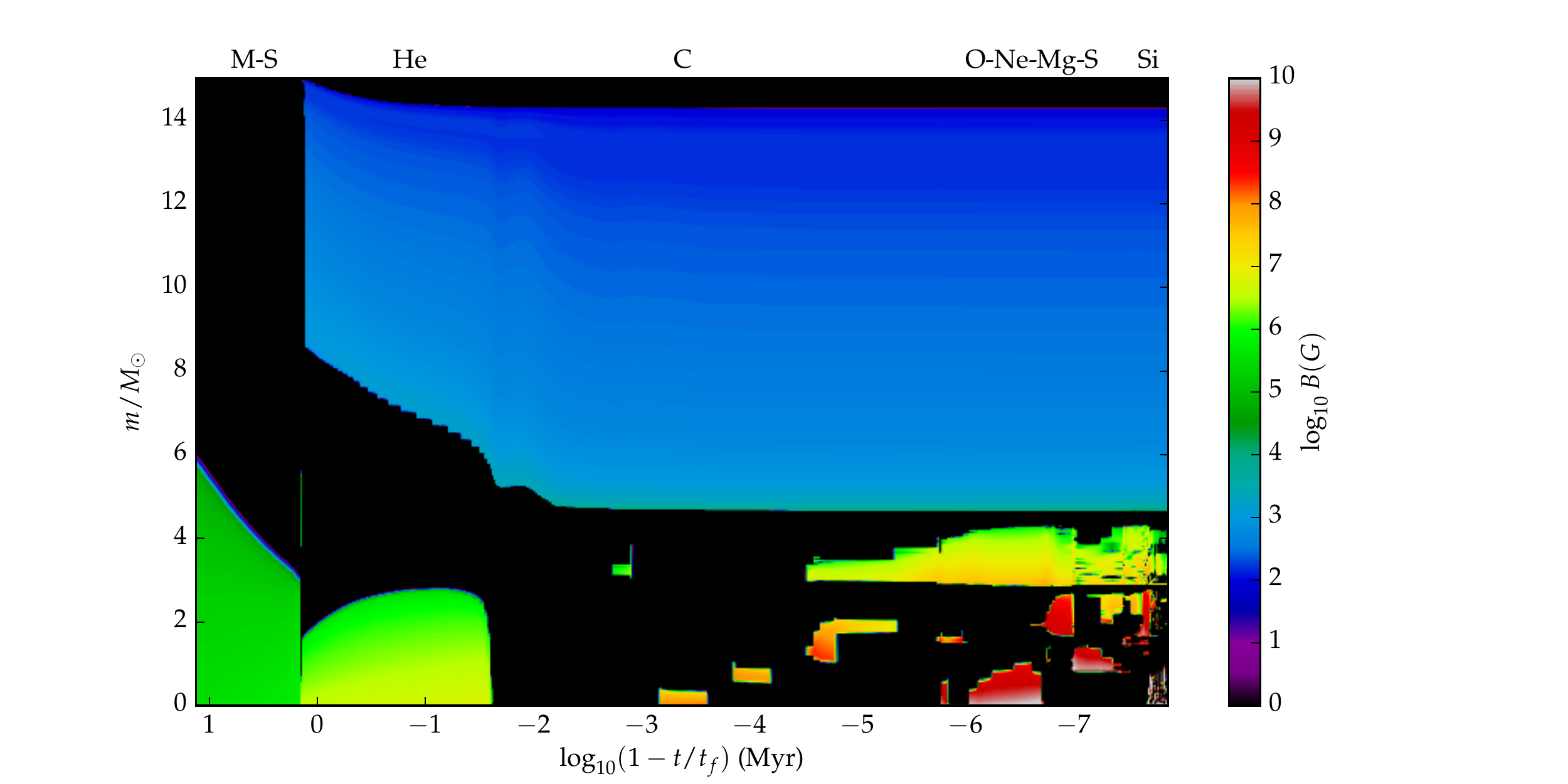}
  \caption{A magnetic Kippenhahn diagram showing the evolution of the equipartition magnetic field for a 15~$M_{\odot}$
    star. The abscissa show the time remaining in Myr before the iron core infall that occurs at $t_f$. The burning
    phase of the core is indicated at the top of the diagram.}
  \label{fig:kipp}
\end{figure}

For a subset of dynamos, like those discussed in \citet{augustson16}, Equation (\ref{eqn:forcescaling}) may hold. Such
dynamos are sensitive to the degree of rotational constraint on the convection and upon the intrinsic ability of the
convection to generate a sustained dynamo.  The inertial term, in particular, may permit a minimum magnetic energy state
to be achieved, bridging the subequipartition slowly rotating dynamos to the rapidly rotating magnetostrophic regime,
where $\mathrm{ME/KE}\propto \mathrm{Ro}^{-1}$. For low Rossby numbers, or large rotation rates, it is possible that the
dynamo can reach superequipartition states where $\mathrm{ME/KE}>1$. Indeed, it may be much greater than unity, as is
expected for the Earth's dynamo \citep[e.g.,][]{roberts13}.

Consider the data for the evolution of a set of MHD simulations using the Anelastic Spherical Harmonic code presented in
\citet{augustson16}. These simulations attempt to capture the dynamics within the convective core of a 10~$M_{\odot}$
B-type star. Given the choices of rotation rates for this suite of simulations, they have nearly three decades of
coverage in Rossby number, as shown in Figure \ref{fig:scaling}(a). In that figure, the force-based scaling given in
Equation (\ref{eqn:forcescaling}) is depicted by the black curve (where $\mathrm{ME/KE}\propto 0.5+\mathrm{Ro}^{-1}$).
This scaling does a reasonable job of describing the nature of the superequipartition state for a given Rossby
number. These simulated convective core dynamos appear to enter a regime of magnetostrophy for the four cases with the
lowest average Rossby number, where the scaling for the magnetostrophic regime is denoted by the dashed blue line in
Figure \ref{fig:scaling}(a).  This transition to the magnetostrophic regime can be better understood through Figure
\ref{fig:scaling}(b), which shows the dynamic Elsasser number
($\Lambda_D = \mathrm{B_{rms}}^2/(8\pi\rho_0\Omega_0\mathrm{v_{rms}}\ell)$, where $\ell$ is the typical length scale of
the current density $\vJ$). So as $\Lambda_D$ approaches unity, the balance between the Lorentz and the Coriolis forces
also approaches unity, indicating that the dynamo is close to magnetostrophy.

The scaling relationship between the magnetic and kinetic energies of convective dynamos in turn provide an estimate of
the rms magnetic field strength in terms of the local rms velocity and density at a particular depth in a convective
zone. Therefore, these relationships permit the construction of magnetic Kippenhahn diagrams that show the equipartition
magnetic field, which is estimated based on the mixing length velocities achieved in stellar evolution models, as shown
in Figure \ref{fig:kipp} for a 15~$M_{\odot}$ star.  During the main sequence, the magnetic field generated by the
dynamo running in the convective core has an estimated rms strength of about $10^6$ Gauss, which is consistent with the
simulations described in \citet{augustson16}.  Likewise, during the helium-burning phase, the equipartition magnetic
field rises to about $10^7$ Gauss.  During subsequent burning phases, the field amplitude continues to rise largely due
to the increasing density of the convective regions where it eventually reaches $10^{10}$ Gauss during the oxygen-neon
and silicon burning stages. The density dependence of the equipartition magnetic field can be seen more directly in the
scaling $ B \propto \rho_{\mathrm{CZ}}^{1/6} L^{1/3}$, which follows from the scaling of the mixing length velocity
discussed above, and by noting the surface luminosity of the star does not change significantly during these late-stage
burning phases.

\section{Conclusions} \label{sec:con}

As discussed in \citet{augustson17a} and \citet{augustson17b}, there appear to be two scaling laws for the level of
equipartition of magnetic and kinetic energies that are applicable to stellar systems, one in the high magnetic Prandtl
number regime and another in the low magnetic Prandtl number regime.  Within the context of the large magnetic Prandtl
number systems, the ratio of the magnetic to the kinetic energy of the system scales as
$\mathrm{ME/KE}\propto\beta + \mathrm{Ro}^{-1}$, where $\beta$ depends upon the details of the non-rotating system, as
mentioned above in \S\ref{sec:forcescaling} and in \citet{augustson16}. For low magnetic Prandtl number and fairly
rapidly rotating systems, such as the geodynamo and rapidly rotating low-mass stars, another scaling relationship may be
more applicable. This scaling relies upon a balance of buoyancy work and magnetic dissipation and it yields a ratio of
magnetic to kinetic energy that scales as the inverse square root of the convective Rossby number
\citep{davidson13,augustson17a}.  In either case, it is likely that the magnetic energy can grow to be near or above
equipartition with the kinetic energy, which allows the estimation of the magnetic energy at various stages of evolution
as shown in Figure \ref{fig:kipp}.  Future magnetic field estimates will consider both the magnetic Prandtl number and
the star's rotational evolution, utilizing angular momentum transport techniques such as those discussed in
\citet{amard16}. Yet, more work is needed to establish more robust scaling relationships that cover a greater range in
both magnetic Prandtl number and Rossby number. Likewise, numerical experiments should explore a larger range of
Reynolds number and level of supercriticality. Indeed, as in \citet{yadav16}, some authors have already attempted to
examine such an increased range of parameters for the geodynamo. Nevertheless, to be more broadly applicable in stellar
physics, there is a need to find scaling relationships that can bridge both the low and high magnetic Prandtl number
regimes that are shown to exist within main-sequence stars. The authors are currently working toward this goal, as will
be presented in an upcoming paper.\\\vspace{-0.1truein}

\noindent \textbf{Acknowledgements} -- K.~C. Augustson and S. Mathis acknowledge support from the ERC SPIRE 647383
grant. A.~S. Brun acknowledges funding by ERC STARS2 207430 grant, INSU/PNST, CNES Solar Orbiter, PLATO and GOLF grants, and
the FP7 SpaceInn 312844 grant. J. Toomre thanks the NASA TCAN grant NNX14AB56G for support.\vspace{-0.2truein}
\bibliography{scaling_cat}

\end{document}